\documentclass[10pt]{article}
\usepackage{natbib}

\title{Magnetic Fields of Nearby Active Galactic Nuclei and
Correlation of the Highest-Energy Cosmic Rays with their
Positions}
\author{M. Yu. Piotrovich, Yu. N. Gnedin, T. M. Natsvlishvili\\
Central Astronomical Observatory at Pulkovo, Saint-Petersburg,
Russia.}

\begin{document}

\maketitle

\begin{abstract}
The correlation between the pointing direction of ultra high
energy cosmic rays and AGN observed by the Pierre Auger
Collaboration is explained in the framework of acceleration
process in AGN. This acceleration process is produced by a
rotating accretion disk around a black hole that is frozen-in
magnetic field. In a result the accretion disk is acting as a
induction accelerator of cosmic rays. We estimate the resulting
magnetic field in the framework of the magnetic coupling process.
The results of our calculations allow to make the conclusion that
the Flat Spectrum Radio Quasars appear the effective cosmic
accelerators. We estimate also the attenuation of highest-energy
cosmic rays in a result of their interaction with ambient
radiation field.

\end{abstract}

\section{Introduction}

Recently, using data collected at the Pierre Auger observatory
during the past 3.7 years, the Pierre Auger Collaboration
\citep{pac07a,pac07b} have demonstrated a correlation between the
arrival directions of cosmic rays with energy above $\sim 6\times
10^{19}$eV and the positions of active galactic nuclei (AGNs)
lying at a distance less than $\sim 75$ Mpc. A confidence level of
anisotropy have been at value of more than 99\% through a number
of special tests \citep{pac07b}. The authors claimed that the
observed correlation is compatible with the hypothesis that cosmic
rays with the highest energies originate from extra-galactic
sources located quite close because their flux is not
significantly attenuated by the Greisen-Zatsepin-Kuz'min effect,
i.e. by interaction with the cosmic background radiation. Though
the present data can't identify AGN as the sources of cosmic rays
unambiguously, AGNs or objects having a similar spatial
distribution can be considered as possible sources. Certainly,
these present data do not identify as the sources of cosmic rays
unambiguously. Therefore, other candidate sources which are
distributed as nearby AGN can't be excluded. Thus,
\citet{dermer07} has considered gamma ray bursts and only blazar
AGN as the sources of UHECRs. More original idea has been
suggested by \citet{grib07}. They suggested the hypothesis that
dark matter consist of superheavy particles with the mass close to
the Grand Unification scale. They claimed that some part of these
particles, which were created from vacuum by the gravitation of
the expanding Universe, can be swallowed by the supermassive black
holes and can decay on visible particles producing the flow of
UHECRs observed by the Auger group.

The interesting look on AGNs as cosmic supercolliders has been
developed by Kardashev and his colleagues
\citep{kardashev95,shatsky03,shatsky02,shatsky06}. They considered
the structure and magnitude of electromagnetic field produced by a
rotating accretion disk around a black hole. The accretion disk
was modeled as a torus filled with plasma and the frozen-in
magnetic field. It acts as an induction accelerator of cosmic rays
along the axis of an accretion disk.

According to \citet{shatsky03} the maximum energy of the
accelerated charged particle is represented as

\begin{equation}
E_{max}\approx 5\times 10^{11}\frac{1}{2\pi^2} \left(\frac{\Omega
R}{c}\right) \left(\frac{\Omega_g r_g}{\Omega R}\right)
\left(\frac{B}{10^4}\right) \left(\frac{M}{M_{\odot}}\right) eV
\label{eq1}
\end{equation}

\noindent where $R$ is the distance from a massive black hole to
an accretion torus, $\Omega$ is an angular velocity of a torus,
$r_g$ is the gravitational radius and $\Omega_g$ is the angular
velocity at $r_g$.

If the basic parameters of the accretion torus near a black hole
are of the same order, i.e. $\Omega\sim\Omega_g$ and $R\sim r_g$,
the Eq.(\ref{eq1}) transforms to

\begin{equation}
E_{max} = 2.5\times 10^{10}\left(\frac{B}{10^4 G}\right)
\left(\frac{M}{M_{\odot}}\right) eV
\label{eq2}
\end{equation}

Another process that can be responsible for generation of ultra
high energy cosmic rays has been most recently considered by
\citet{koide08}. They considered the magnetic reconnection as a
basic process of energy extraction from a rotating supermassive
black hole. Magnetic reconnection is one of the important
processes for the particle acceleration and heating. The
relativistic reconnection can be also responsible for generation
of ultra high energy cosmic rays. Genuinely this process occurs if
the magnetic energy exceeds the plasma energy including its rest
energy. And in this case the magnetic field of AGN takes a key
role in the process of acceleration and generation of cosmic rays.

The basic goal of our paper is to obtain the real estimation of
the magnetic field strength in the close vicinity of accreting
supermassive black hole (SMBH) in AGNs and to obtain the more real
estimations of $E_{max}$ from Eq.(\ref{eq2}).

\section{Magnetic coupling process and determination of magnetic
field\\ strengths of supermassive black holes (SMBH)}

Recently \citet{li02,wang02,wang03,zhang05}, have studied as a
possible mechanism for transferring energy and angular momentum
from rotating black hole to its surrounding disk the magnetic
coupling process through the closed field lines connecting the
Kerr black hole with its accretion disk. This process can be
considered as one of the variants of the Blanford-Znajek (BZ)
process \citep{blanford77}. It is assumed that the disk is stable,
perfectly conducting, thin and Keplerian. The magnetic field is
assumed to be constant on the black hole horizon and to vary as a
power law with the radial coordinate of the accretion disk.

Since the magnetic field on the horizon $B_H$ is brought and held
by its surrounding magnetized matter of a disk there must exist
some relation between the magnetic field strength and accretion
disk and finally the bolometric luminosity of AGN. This relation
have been obtained by \citet{gan07}. It has a form:

\begin{equation}
B_H = k \frac{(2 L_{bol} / \varepsilon
c)^{\frac{1}{2}}}{R_H},\,\,\, k\approx 1
\label{eq3}
\end{equation}

\noindent Here $L_{bol} = \varepsilon \dot{M} c^2$, $R_{H} =
\frac{G M}{c^2} \left[1 + \sqrt{1 - (a/M)^2}\right]$, $\dot{M}$ is
the accretion rate, $\varepsilon$ is the radiation conversion
efficiency (calculated, for example, by
\citet{novikov73,krolik07,shapiro07}). The Eq.(\ref{eq2}) is
transformed into

\begin{equation}
B_H = 6.2\times 10^8
\left(\frac{M_{\odot}}{M_{BH}}\right)^{\frac{1}{2}}
\left(\frac{\eta}{\varepsilon}\right)^{\frac{1}{2}} \frac{1}{1 +
\sqrt{1 - \left(\frac{a}{M_{BH}}\right)^2}}
\label{eq4}
\end{equation}

\noindent where $(a/M_{BH})$ is the Kerr parameter and $\eta =
L_{bol} / L_{Edd}$ and $L_{Edd} = 1.3\times 10^{38} (M_{BH} /
M_{\odot})$ is the Eddington luminosity.

The relation (\ref{eq4}) can be used for estimation $E_{max}$ from
Eq.(\ref{eq2}). After substituting Eq.(\ref{eq4}) into
Eq.(\ref{eq2}) one gets:

\begin{equation}
E_{max}\approx 1.6\times 10^{15}
\left(\frac{M_{BH}}{M_{\odot}}\right)^{\frac{1}{2}}
\left(\frac{\eta}{\varepsilon}\right)^{\frac{1}{2}} \frac{1}{1 +
\sqrt{1 - \left(\frac{a}{M_{BH}}\right)^2}}
\label{eq5}
\end{equation}

\section{Estimating maximal energy for specific active galactic
nuclei (AGN)}

We are starting with commonly accepted parameters for accreting
Kerr supermassive black holes (SMBHs).

If one chooses for Eq.(\ref{eq5}) the next values of physical
parameters: $M_{BH} = 10^9 M_{\odot}$, $\eta\approx 1$,
$\varepsilon\approx 0.1$ and $(a / M_{BH})\approx 1.0$ one gets

\begin{equation}
E_{max} = 1.6\times 10^{20} eV
\label{eq6}
\end{equation}

Below we consider more real situation. A systematic analysis of a
large sample of radio-loud AGN available in the BeppoSAX public
archive has been performed by \citet{grandi05}. Their sample
includes AGN of various types: Narrow Line Radio Galaxies (NLRG),
Broad Line Radio Galaxies (BLRG), Steep Spectrum Radio Quasars
(SSRQ) and Flat Spectrum Radio Quasars (FSRQ) (see Table 6 from
their paper).

Our calculations of the maximal energy magnitude for these types
of AGN are presented in Tables 1, 2 and 3. One can see from these
results that the maximal energy value essentially exceeding the
Greisen-Zatsepin-Kuz'min (GZK) cutoff is reached only by FSRQ
sources. It is well-known that this class is characterized by the
presence of a non-thermal beamed component. It shows also a
featureless continuum and a significantly flatter average spectral
slope. These facts give evidence that the most effective
acceleration process takes place in this type of quasars.

\begin{table*}
\centering
\begin{minipage}{180mm}
\caption{Magnetic field strength.}
\begin{tabular}{@{}lccrrrrrr}
\hline

Source & $\log{\left(\frac{M_{BH}}{M_{\odot}}\right)}$ &
$\log{\left(\frac{L_{bol}}{L_{Edd}}\right)}$ & $B_H [G]$ & $B_H
[G]$ & $B_H [G]$ & $B_H [G]$ & $B_H [G]$ & $B_H [G]$ \\

 & & & ($\frac{a}{M} = 0.0$   & ($\frac{a}{M} = 0.5$   & ($\frac{a}{M} = 0.9$  & ($\frac{a}{M} = 0.95$ & ($\frac{a}{M} = 0.998$ & ($\frac{a}{M} = 1.0$ \\

 & & & $\varepsilon = 0.057$) & $\varepsilon = 0.081$) & $\varepsilon = 0.16$) & $\varepsilon = 0.19$) & $\varepsilon = 0.32$)  & $\varepsilon = 0.42$) \\

\hline

BLRG      & 8.64 & -1.59 & $10^4$           & $9\times 10^3$   &
$8.3\times 10^3$ & $5.6\times 10^3$ & $5.3\times 10^3$ & $5\times
10^3$ \\

NLRG      & 8.14 & -2.92 & $3.8\times 10^3$ & $3.4\times 10^3$ &
$3.2\times 10^3$ & $3.2\times 10^3$ & $3\times 10^3$   &
$2.8\times 10^3$ \\

SSRQ      & 9.29 & -0.90 & $10^4$           & $9.4\times 10^3$ &
$8.7\times 10^3$ & $8.7\times 10^3$ & $8.3\times 10^3$ &
$7.7\times 10^3$ \\

FSRQ      & 9.01 &  0.44 & $6.7\times 10^4$ & $6.1\times 10^4$ &
$5.6\times 10^4$ & $5.6\times 10^4$ & $5.4\times 10^4$ & $5\times
10^4$ \\

Seifert 1 & 7.23 & -0.59 & $1.6\times 10^5$ & $1.4\times 10^5$ &
$1.3\times 10^5$ & $1.3\times 10^5$ & $1.3\times 10^5$ &
$1.2\times 10^5$ \\

\hline
\end{tabular}
\end{minipage}
\end{table*}

\begin{table*}
\centering
\begin{minipage}{180mm}
\caption{Maximal energy of cosmic rays.}
\begin{tabular}{@{}lccrrrrrr}
\hline

Source & $\log{\left(\frac{M_{BH}}{M_{\odot}}\right)}$ &
$\log{\left(\frac{L_{bol}}{L_{Edd}}\right)}$ & $E_{max} [eV]$ &
$E_{max} [eV]$ & $E_{max} [eV]$ & $E_{max} [eV]$ & $E_{max} [eV]$
& $E_{max} [eV]$ \\

 & & & ($\frac{a}{M} = 0.0$   & ($\frac{a}{M} = 0.5$   & ($\frac{a}{M} = 0.9$  & ($\frac{a}{M} = 0.95$ & ($\frac{a}{M} = 0.998$ & ($\frac{a}{M} = 1.0$ \\

 & & & $\varepsilon = 0.057$) & $\varepsilon = 0.081$) & $\varepsilon = 0.16$) & $\varepsilon = 0.19$) & $\varepsilon = 0.32$)  & $\varepsilon = 0.42$) \\

\hline

BLRG      & 8.64 & -1.59 & $10^{19}$           & $9.8\times
10^{18}$ & $9\times 10^{18}$   & $9\times 10^{18}$   & $8.6\times
10^{18}$  & $8\times 10^{18}$ \\

NLRG      & 8.14 & -2.92 & $1.3\times 10^{18}$ & $1.2\times
10^{18}$  & $10^{18}$           & $10^{18}$           & $10^{18}$
& $9\times 10^{17}$ \\

SSRQ      & 9.29 & -0.90 & $5\times 10^{19}$   & $4.6\times
10^{19}$  & $4.2\times 10^{19}$ & $4.2\times 10^{19}$ & $4\times
10^{19}$    & $3.8\times 10^{19}$ \\

FSRQ      & 9.01 &  0.44 & $1.7\times 10^{20}$ & $1.6\times
10^{20}$  & $1.4\times 10^{20}$ & $1.4\times 10^{20}$ &
$1.37\times 10^{20}$ & $1.3\times 10^{20}$ \\

Seifert 1 & 7.23 & -0.59 & $7\times 10^{18}$   & $6\times 10^{18}$
& $5.7\times 10^{18}$ & $5.7\times 10^{18}$ & $5.4\times 10^{18}$
& $5\times 10^{18}$ \\

\hline
\end{tabular}
\end{minipage}
\end{table*}

\begin{table*}
\centering \caption{Retrograde motion ($\frac{a}{M} = -0.9$,
$\varepsilon = 0.039$).}
\begin{tabular}{@{}lrr}
\hline

Source & $B_H [G]$ & $E_{max} [eV]$ \\

\hline

BLRG      & $1.7\times 10^4$  & $1.8\times 10^{19}$ \\

NLRG      & $6.5\times 10^3$  & $2.24\times 10^{18}$ \\

SSRQ      & $1.76\times 10^4$ & $8.6\times 10^{19}$ \\

FSRQ      & $1.6\times 10^5$  & $4.2\times 10^{20}$ \\

Seifert 1 & $2.7\times 10^5$  & $1.2\times 10^{19}$ \\

\hline
\end{tabular}
\end{table*}

\section{The interaction of ultra high energy cosmic rays with
intrinsic radiation of active galactic nuclei and quasars}

The basic problem for the mechanism considered here is the
attenuation of the cosmic ray flux above $\sim 5\times 10^{19}$eV
in a result of interaction of this flux with the ambient radiation
field around of a supermassive black hole. It is well known that,
namely, this attenuation by extragalactic background radiation
produces the so-called GZK effect. The similar effect by ambient
radiation of AGN or QSO should be estimated in our situation.

We estimate of the optical thickness $\tau$ respect to the
interaction of the cosmic ray flux with the ambient radiation of
AGN and QSO. The value of the optical thickness respect to the
$p\gamma$ process is

\begin{equation}
\tau\approx \frac{L\sigma}{4\pi c R k_B T}
\label{eq7}
\end{equation}

\noindent where $L$ is the radiation luminosity, $R$ is the
characteristic scale distance of the radiation field, $T$ is the
temperature of a radiation field, $\sigma$ is the cross-section of
the interaction of cosmic protons with the radiation field that is
$\sigma\approx 10^{-28} cm^2$ (\citet{atoyan03}, see also
\citet{reynoso08}).

The temperature $T$ is determined by the temperature of an
accretion disk $T_e$ and it is determined by \citep{bonning07}:

\begin{equation}
T_e = T_0 \left(\frac{M_{\odot}}{M_{BH}}\right)^{1/4}
\left(\frac{L}{L_{Ed}}\right)^{1/4}
\left(\frac{R_{in}}{R}\right)^{3/4}
\label{eq8}
\end{equation}

\noindent where $R_{in}$ is the radius of an accretion disk.

The relation (\ref{eq8}) corresponds to the standard accretion
model of Shakura and Sunyaev. The value of $T_0$ is determined at
the radius $R_{in}$ that is closed to the event horizon and
corresponds to X-ray temperature, i.e. $T_0\approx 10^8$K.

We consider two situations: (a) interaction of the cosmic ray flux
with X-ray radiation field and (b) with optical and UV radiation
field.

The radiation flux is falling down with a distance as $R^{-2}$. We
determine the characteristic scale distance for X-ray radiation
field as $R \approx 3 R_g$, where $R_g$ is the Schwarzschild
radius. For the optical and UV-radiation the characteristic
scale-distance is $\sim (10^2\div 10^3) R_g$. In a result we
obtain:

\begin{equation}
\tau \approx 12.8 \left(\frac{L}{L_{Ed}}\right)^{3/4}
\left(\frac{M_{BH}}{M_{\odot}}\right)^{1/4} \frac{1}{r^{1/4}};
\,\, R = r R_g \label{eq9}
\end{equation}

This Eq.(\ref{eq9}) means that $\tau\sim
(M_{\odot}/M_{BH})^{1/2}$, i.e. the increase of the mass of SMBH
decreases the optical thickness.

For $M_{BH} = 10^9 M_{\odot}$ and typical values $L = L_x =
10^{43}$erg/s and $L_{Opt,UV}\approx 10^{44}$erg/s we obtain from
Eq.(\ref{eq9}): $\tau = \tau_x \approx 1.2$ and $\tau =
\tau_{Opt,UV} \approx 6$. These values demonstrate that the escape
of a part of accelerated protons of UHE from the nearest
environment of SMBH is very likely. It should be noted that the
cosmic ray luminosity of the active galactic nuclei constitutes a
fraction $\sim 10^{-4}$ of the optical one \citep{dedenko08}. This
result does not contradict to our estimations.

\section{Conclusions}

The correlation between the pointing directions of ultra high
energy cosmic rays observed by the High Resolution Fly's Eye
experiment and Active Galactic Nuclei visible from its northern
hemisphere location can be explained in the framework of
acceleration process in AGN as cosmic supercollider.
\citet{kardashev95,shatsky02,shatsky06} suggested that this
acceleration process produced by a rotating accretion disk around
a black hole that is frozen-in magnetic field. This accretion disk
is acting as an induction accelerator of cosmic rays.

We estimated the resulting magnetic field in the framework of the
magnetic coupling process. The results of our calculation allow to
make the conclusion that only the Flat Spectrum Radio Quasars can
be revealed as the effective cosmic accelerators. Only for these
objects the maximal energy of cosmic rays can essentially exceed
the GZK cutoff.

Our conclusions are not totally exclude another probable scenario
of UHECR generation. We mean the process of generation of cosmic
ray particles with ultra-high energy in a result of acceleration
in jets and extended envelopes surrounding AGN. This scenario was
recently developed by a number of authors
\citep{dermer08,allard08,berezhko08,bierman08}. The decisive
choice between various probable scenarios will be done in the
result of future cosmic observations and move details confirmation
of the Pierre Auger Collaboration data.

\section*{Acknowledgments}

Our research was supported by the RFBR (project No. 07-02-00535a),
Program of Prezidium of RAS ''Origin and Evolution of Stars and
Galaxies'', the Program of the Department of Physical Sciences of
RAS ''Extended Objects in the Universe''.

This research was also supported by the Grant of President of
Russian Federation ''The Basic Scientific Schools''
NS-6110.2008.2.

M.Yu. Piotrovich acknowledges the Council of Grants of President
of Russian Federation for Young Scientists, grant No. 4101.2008.2.

We would like to thank academician N.S. Kardashev for very useful
comments.


\begin{thebibliography}{99}
\bibitem[\protect\citeauthoryear{Allard et al.}{2008}]{allard08} Allard D., Busca N.G., Decerpit G. et al., 2008, arXiv:0805.4779
\bibitem[\protect\citeauthoryear{Atoyan \& Dermer}{2003}]{atoyan03} Atoyan A.M., Dermer C.D., 2003, ApJ, 586, 79
\bibitem[\protect\citeauthoryear{Berezhko}{2008}]{berezhko08} Berezhko E.G., 2008, arXiv:0809.0734
\bibitem[\protect\citeauthoryear{Bierman et al.}{2008}]{bierman08} Bierman P.L., Becker J.K., Caramete L. et al., 2008, arXiv:0811.1848
\bibitem[\protect\citeauthoryear{Blanford \& Znajek}{1977}]{blanford77} Blanford R.D., Znajek R.L., 1977, MNRAS, 179, 433
\bibitem[\protect\citeauthoryear{Bonning et al.}{2007}]{bonning07} Bonning E.W., Cheng L., Shields G.A., Salviander S., Gebhardt K., 2007, ApJ, 659, 211
\bibitem[\protect\citeauthoryear{Collin}{2008}]{collin08} Collin S., 2008, arXiv:0811.1731
\bibitem[\protect\citeauthoryear{Dedenko et al.}{2008}]{dedenko08} Dedenko L.G., Podgrudkov D.A., Roganova T.M., Fedorova G.F., 2008, arXiv:0811.0722
\bibitem[\protect\citeauthoryear{Dermer}{2007}]{dermer07} Dermer C.D., 2007, arXiv:0711.2804
\bibitem[\protect\citeauthoryear{Dermer et al.}{2008}]{dermer08} Dermer C.D., Razzaque S., Finke J.D., Atoyan A., 2008, arXiv:0811.1160
\bibitem[\protect\citeauthoryear{Gan, Wang \& Li}{2007}]{gan07} Gan Z.-M., Wang D.-X., Li Y., 2007, astro-ph/0701532
\bibitem[\protect\citeauthoryear{Grandi, Malaguti \& Fiocchi}{2005}]{grandi05} Grandi P., Malaguti G., Fiocchi M., 2005, astro-ph/0511784
\bibitem[\protect\citeauthoryear{Grib \& Pavlov}{2007}]{grib07} Grib A.A., Pavlov Yu.V., 2007, arXiv:0712.2667
\bibitem[\protect\citeauthoryear{Kardashev}{1995}]{kardashev95} Kardashev N.S., 1995, MNRAS, 276, 515
\bibitem[\protect\citeauthoryear{Koide \& Arai}{2008}]{koide08} Koide S., Arai K., 2008, arXiv:0805.0044
\bibitem[\protect\citeauthoryear{Krolik}{2007}]{krolik07} Krolik J.H., 2007, arXiv:0709.1489
\bibitem[\protect\citeauthoryear{Li}{2002}]{li02} Li L.-X., 2002, ApJ, 567, 463
\bibitem[\protect\citeauthoryear{Novikov \& Thorne}{1973}]{novikov73} Novikov I.D., Thorne K.S., 1973, Astrophysics of black holes, Black Holes (Les astres occlus)
\bibitem[\protect\citeauthoryear{Reynoso \& Romero}{2008}]{reynoso08} Reynoso M.M., Romero G.E., 2008, arXiv:0811.1383
\bibitem[\protect\citeauthoryear{Shapiro}{2007}]{shapiro07} Shapiro S.L., 2007, arXiv:0711.1537
\bibitem[\protect\citeauthoryear{Shatsky \& Kardashev}{2002}]{shatsky02} Shatsky A.A., Kardashev N.S., 2002, Astrom.rep., 46, 639
\bibitem[\protect\citeauthoryear{Shatsky \& Kardashev}{2006}]{shatsky06} Shatsky A.A., Kardashev N.S., 2006, astro-ph/0209465
\bibitem[\protect\citeauthoryear{Shatsky}{2003}]{shatsky03} Shatsky A.A., 2003, astro-ph/0301535
\bibitem[\protect\citeauthoryear{The Pierre Auger Collaboration}{2007a}]{pac07a} The Pierre Auger Collaboration, 2007a, arXiv:0711.2256
\bibitem[\protect\citeauthoryear{The Pierre Auger Collaboration}{2007b}]{pac07b} The Pierre Auger Collaboration, 2007b, arXiv:0712.2843
\bibitem[\protect\citeauthoryear{Wang et al.}{2003}]{wang03} Wang D.-X., Ma R.-Y., Lei W.-H., Yao C.-Z., 2003, ApJ, 595, 109
\bibitem[\protect\citeauthoryear{Wang \& Xiao}{2002}]{wang02} Wang D.-X., Xiao K., 2002, MNRAS, 335, 655
\bibitem[\protect\citeauthoryear{Zhang, Lu \& Zhang}{2005}]{zhang05} Zhang W.M., Lu Y., Zhang S.N., 2005, astro-ph/0501365
\end{thebibliography}
\end{document}